\documentclass{sig-alternate}

\usepackage[utf8]{inputenc}
\usepackage[T1]{fontenc}
\usepackage{textcomp}
\usepackage[french, english]{babel}
\usepackage{graphicx}
\usepackage{url}
\usepackage{backnaur}
\usepackage{tikz}
\usepackage{xcolor}
\usepackage{color}
\usepackage{listings}
\usepackage{inconsolata}

\usepackage[hyperref=true,%
            url=false,%
            isbn=false,%
            style=numeric,%
            maxcitenames=3,%
            maxbibnames=100,%
            backend=bibtex,%
            block=none]{biblatex}
\bibliography{references.bib}

\definecolor{red}{rgb}{1,0,0.29}
\definecolor{green}{rgb}{0,0.91,0.75}
\definecolor{lightgray}{rgb}{.9,.9,.9}
\definecolor{gray}{rgb}{.6,.6,.6}
\definecolor{darkgray}{rgb}{.4,.4,.4}

\newcommand{\CodeSymbol}[1]{
    \bfseries\textcolor{red}{#1}
}

\lstdefinelanguage{js}{
  keywords={break, case, catch, continue, debugger, default, delete, do, else, false, finally, for, function, if, in, instanceof, new, null, return, switch, this, throw, true, try, typeof, var, void, while, with},
  morecomment=[l]{//},
  morecomment=[s]{/*}{*/},
  morestring=[b]',
  morestring=[b]",
  ndkeywords={class, export, boolean, throw, implements, import, this},
  keywordstyle=\color{red}\bfseries,
  ndkeywordstyle=\color{red}\bfseries,
  identifierstyle=\color{black},
  commentstyle=\color{lightgray}\ttfamily,
  stringstyle=\color{gray}\bfseries,
  basicstyle=\color{gray}\ttfamily\scriptsize,
  sensitive=true,
  escapeinside={\/\/\@}{\@}
}

\lstdefinelanguage{flx}{
  keywords={flx, fluxion},
  morecomment=[l]{//},
  morecomment=[s]{/*}{*/},
  morestring=[b]',
  morestring=[b]",
  literate= {+>}{{\ttfamily\scriptsize>>}}2
            {->}{{\CodeSymbol{->}}}2
            {>>}{{\CodeSymbol{>>}}}2, % This is a hack to get -> and >> be highlighted like keywords.
  ndkeywords={main, express\_dispatcher, app\_js, get, handler, reply, anonymous\_1000, null},
  keywordstyle=\color{red}\bfseries,
  ndkeywordstyle=\color{black}\bfseries,
  identifierstyle=\color{black},
  commentstyle=\color{lightgray}\ttfamily,
  stringstyle=\color{gray}\bfseries,
  basicstyle=\color{gray}\ttfamily\scriptsize,
  sensitive=true,
  escapeinside={\/\/\@}{\@}
}

\newcommand{\userlstset}[1]{
  \lstset{ %
    numberstyle=\tiny,               % the size of the fonts that are used for the line-numbers
    numbers=left,                    % where to put the line-numbers
    stepnumber=1,                    % the step between two line-numbers. If it is 1 each line will be numbered
    numbersep=5pt,                   % how far the line-numbers are from the code
    showspaces=false,                % show spaces adding particular underscores
    showstringspaces=false,          % underline spaces within strings
    showtabs=false,                  % show tabs within strings adding particular underscores
    tabsize=2,                       % sets default tabsize to 2 spaces
    captionpos=b,                    % sets the caption-position to bottom
    breaklines=true,                 % sets automatic line breaking
    breakautoindent = true,          %
    breakatwhitespace=false,         % sets if automatic breaks should only happen at whitespace
    escapeinside={\@}{\@},           % if you want to add a comment within your code
    language=#1,                     % choose the language of the code
  } %
}

\newcommand{\ic}[1]{\lstinline|#1|}

\lstnewenvironment{code}[1][js]{%
  \userlstset{#1}%
}{%
}

\newcommand{\ftnt}[1]{%
\footnote{\small{\url{#1}}}%
}

\newcommand*\circled[1]{%
  \tikz[baseline=(char.base)]{%
    \node[shape=circle,draw,inner sep=0.8pt] (char) {#1};%
  }%
}

\begin{document}

\title{
  Transforming Javascript Event-Loop Into a Pipeline
}

\numberofauthors{2}
\author{
\alignauthor
Etienne Brodu, Stéphane Frénot\\
  \email{\textsf{\normalsize{\{etienne.brodu, stephane.frenot\}@insa-lyon.fr}}}\\
  \affaddr{\textsf{\small{Univ Lyon, INSA Lyon, Inria, CITI, F-69621 Villeurbanne, France}}}
\and
\alignauthor
Frédéric Oblé\\
  \email{\textsf{\normalsize{frederic.oble@worldline.com}}}\\
  \affaddr{\textsf{\small{Worldline, Bât. Le Mirage, 53 avenue Paul Krüger}}}\\
  \affaddr{\textsf{\small{CS 60195, 69624 Villeurbanne Cedex}}}\\
}

% \CopyrightYear{2016}
% \setcopyright{licensedothergov}
% \conferenceinfo{SAC 2016,}{April 04 - 08, 2016, Pisa, Italy}
% \isbn{978-1-4503-3739-7/16/04}\acmPrice{\$15.00}
% \doi{http://dx.doi.org/10.1145/2851613.2851745}

\maketitle

\begin{abstract}

The development of a real-time web application often starts with a feature-driven approach allowing to quickly react to users feedbacks.
However, this approach poorly scales in performance.
Yet, the user-base can increase by an order of magnitude in a matter of hours. This first approach is unable to deal with the highest connections spikes.
It leads the development team to shift to a scalable approach often linked to new development paradigm such as dataflow programming.
This shift of technology is disruptive and continuity-threatening.
To avoid it, we propose to abstract the feature-driven development into a more scalable high-level language.
Indeed, reasoning on this high-level language allows to dynamically cope with user-base size evolutions.

We propose a compilation approach that transforms a Javascript, single-threaded real-time web application into a network of small independent parts communicating by message streams.
We named these parts \textit{fluxions}, by contraction between a flow\footnote{flux in french} and a function.
The independence of these parts allows their execution to be parallel, and to organize an application on several processors to cope with its load, in a similar way network routers do with IP traffic.
We test this approach by applying the compiler to a real web application.
We transform this application to parallelize the execution of an independent part and present the result.

\end{abstract}

\category{Software and its engineering}{Software notations and tools}{Compilers}[Runtime environments]
\terms{Compilation, dataflow, code transformation}
\keywords{Flow programming, Web, Javascript}

\eject

\section{Introduction}

\textit{``Release early, release often''}, \textit{``Fail fast''}.
The growth of a real-time web service is partially due to Internet's capacity to allow very quick releases of a minimal viable product (MVP).
It is crucial for the prosperity of such project to quickly validate that it meets the needs of its users.
Indeed, misidentifying the market needs is the first reason for startup failure\ftnt{https://www.cbinsights.com/blog/startup-failure-post-mortem/}.
Hence the development team quickly concretizes an MVP using a feature-driven approach and iterates on it.

The service needs to be scalable to be able to respond to the growth of its user-base.
However, feature-driven development best practices are hardly compatible with the required parallelism.
The features are organized in modules which disturb the organization of a parallel execution \cite{Clements2013a,Hughes1989,Parnas1972}.
Eventually the growth requires to discard the initial approach to adopt a more efficient processing model.
Many of the most efficient models decompose applications into execution units \cite{Fox1997, Welsh2000, Dean2008}.
However, these tools are in disruption from the initial approach.
This shift causes the development team to spend development resources in background to start over the initial code base, without adding visible value for the users.
It is a risk for the evolution of the project.
Running out of cash and missing the right competences are the second and third reasons for startup failures$^2$.

The risk described above comes from a disruption between the two levels of application expression, the feature level and the execution level.
To avoid this risk and allow a continuous development process, we propose a tool to automatically map one level onto the other, and make the transition.

We focus on web applications driven by users requests and developed in Javascript using the \textit{Node.js}\ftnt{https://nodejs.org/} execution environment.
Javascript is widely adopted\ftnt{http://githut.info/}\ftnt{http://stackoverflow.com/tags} to develop web applications, and its event-loop model is very similar to a pipeline architecture.
So we propose a compiler to transform an application into a pipeline of parallel stages communicating by message streams.
We named these stages \textit{fluxions}, by contraction between a flux and a function.

We present a proof of concept for this compilation approach.
Section \ref{section:model} describes the execution environment targeted by this compiler.
Then, section \ref{section:compiler} presents the compiler, and section \ref{section:evaluation} its evaluation.
Section \ref{section:related} compare our work with related works.
And finally, we conclude this paper.
\section{Fluxional execution model} \label{section:model}

This section presents an execution model to provide scalability to web applications with a granularity of parallelism at the function level.
Functions are encapsulated in autonomous execution containers with their state, so as to be mobile and parallel, similarly to the actors model.
The communications are similar to the dataflow programming model, which allows to reason on the throughput of these streams, and to react to load increases \cite{Bartenstein2014}.

The fluxional execution model executes programs written in our high-level fluxionnal language, whose grammar is presented in figure \ref{fig:flx-lang}.
An application $\bnfpn{program}$ is partitioned into parts encapsulated in autonomous execution containers named \textit{fluxions} $\bnfpn{flx}$.
The following paragraphs present the \textit{fluxions} and the messaging system to carry the communications between \textit{fluxions}, and then an example application using this execution model.

\subsection{Fluxions and Messaging System}

A \textit{fluxion} $\bnfpn{flx}$ is named by a unique identifier $\bnfpn{id}$ to receive messages, and might be part of one or more groups indicated by tags $\bnfpn{tags}$.
A \textit{fluxion} is composed of a processing function $\bnfpn{fn}$, and a local memory called a \textit{context} $\bnfpn{ctx}$.

At a message reception, the \textit{fluxion} modifies its \textit{context}, and sends messages to downstream \textit{fluxions} on its output streams $\bnfpn{streams}$.
The \textit{context} stores the state on which a \textit{fluxion} relies between two message receptions.
The messaging system queues the output messages for the event loop to process them later by calling the downstream \textit{fluxions}.

In addition to message passing, the execution model allows \textit{fluxions} to communicate by sharing state between their \textit{contexts}.
The fluxions that need this synchronization are grouped with the same tag, and loose their independence.

There are two types of streams, \textit{start} and \textit{post}, which correspond to the nature of the rupture point producing the stream.
A variable created within a chain of \textit{post} streams requires more synchronization than a variable created upstream a \textit{start} stream.
The two types and implications of rupture points are further detailed in section \ref{section:compiler}.
\textit{Start} rupture points are indicated with a double arrow ($\to$ \hspace{-1.4em} $\to$ or \texttt{>>}) and \textit{post} rupture points with a simple arrow ($\to$ or \texttt{->}).

\begin{figure}[h]
\vspace{-0.6\baselineskip}
\begin{bnf*}
  \bnfprod{program}    {\bnfpn{flx} \bnfor \bnfpn{flx} \bnfsp \bnftd{eol} \bnfsp \bnfpn{program}}\\
  \bnfprod{flx}        {\bnfts{\texttt{flx}} \bnfsp \bnfpn{id} \bnfsp \bnfpn{tags} \bnfsp \bnfpn{ctx} \bnfsp \bnftd{eol} \bnfsp \bnfpn{streams} \bnfsp \bnftd{eol} \bnfsp \bnfpn{fn}}\\
  \bnfprod{tags}       {\bnfts{\texttt{\&}} \bnfsp \bnfpn{list} \bnfor \bnftd{empty string}}\\
  \bnfprod{streams}    {\bnfts{\texttt{null}} \bnfor \bnfpn{stream} \bnfor \bnfpn{stream} \bnfsp \bnftd{eol} \bnfsp \bnfpn{streams}}\\
  \bnfprod{stream}     {\bnfpn{type} \bnfsp \bnfpn{dest} \bnfsp [\bnfpn{msg}]}\\
  \bnfprod{dest}       {\bnfpn{list}}\\
  \bnfprod{ctx}        {\bnfts{\texttt{\{}} \bnfpn{list} \bnfts{\texttt{\}}}}\\
  \bnfprod{msg}        {\bnfts{\texttt{[}} \bnfpn{list} \bnfts{\texttt{]}}}\\
  \bnfprod{list}       {\bnfpn{id} \bnfor \bnfpn{id} \bnfsp \bnfts{,} \bnfsp \bnfpn{list}}\\
  \bnfprod{type}       {\bnfts{\texttt{>}\texttt{>}} \bnfor \bnfts{\texttt{-}\texttt{>}}}\\
  \bnfprod{id}         {\bnftd{Identifier}}\\
  \bnfprod{fn}         {\bnftd{Source language with~} \bnfpn{stream} \bnftd{~placeholders}}\\
\end{bnf*}
\vspace{-2.5\baselineskip}
\caption{Syntax of a high-level language to represent a program in the fluxionnal form}
\label{fig:flx-lang}
\end{figure}

\subsection{Example}

\begin{code}[js,
  caption={Example web application},
  label={lst:source}]
var app = require('express')(),
    fs = require('fs'),
    count = 0; //@\label{lst:source-counter}@

app.get('/', function handler(req, res){ //@\label{lst:source-handler}@
  fs.readFile(__filename, function reply(err, data) {
    count += 1;
    res.send(err || template(count, data)); //@\label{lst:source-send}@
  });
}); //@\label{lst:source-handler-end}@

app.listen(8080);
\end{code}

The fluxional execution model is illustrated with an example application presented in listing \ref{lst:source}.
This application reads a file, and sends it back along with a request counter.
The \texttt{handler} function, line \ref{lst:source-handler} to \ref{lst:source-handler-end}, receives the input stream of requests.
The \texttt{count} variable at line \ref{lst:source-counter} counts the requests, and needs to be saved between two messages receptions.
The \texttt{template} function formats the output stream to be sent back to the client.
The \texttt{app.get} and \texttt{res.send} functions, lines \ref{lst:source-handler} and \ref{lst:source-send}, interface the application with the clients.
Between these two interface functions is a chain of three functions to process the client requests : \texttt{app.get} $\to$ \hspace{-1.4em} $\to$ \texttt{handler} $\to$ \texttt{reply}.
This chain of functions is transformed into a pipeline, expressed in the high-level fluxionnal language in listing \ref{lst:fluxional}.
The transformation process between the source and the fluxional code is explained in section \ref{section:compiler}.

\begin{figure}[h!]
  \includegraphics[width=\linewidth]{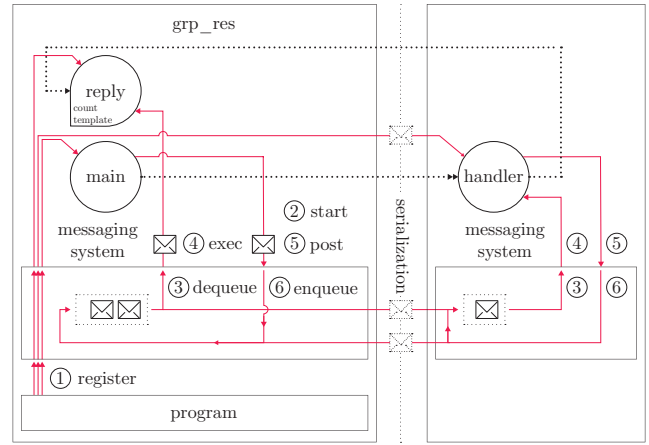}
  \caption{The fluxionnal execution model in details}
  \label{fig:MesSys}
\end{figure}

The execution is illustrated in figure \ref{fig:MesSys}.
The dashed arrows between fluxions represent the message streams as seen in the fluxionnal application.
The plain arrows represent the operations of the messaging system during the execution.
These steps are indicated by numeroted circles.
The \textit{program} registers its fluxions in the messageing system, \circled{1}.
The fluxion \textit{reply} has a context containing the variable \texttt{count} and \texttt{tem\-plate}.
When the application receives a request, the first fluxion in the stream, \textit{main}, queues a \texttt{start} message containing the request, \circled{2}.
This first message is to be received by the next fluxion \textit{handler}, \circled{3}, and triggers its execution, \circled{4}.
The fluxion \textit{handler} sends back a message, \circled{5}, to be enqueued, \circled{6}.
The system loops through steps \circled{3} through \circled{6} until the queue is empty.
This cycle starts again for each new incoming request causing another \texttt{start} message.

\begin{code}[flx, caption={Example application expressed in the high-level fluxional language}, label={lst:fluxional}]
flx main & grp_res
>> handler [res]
  var app = require('express')(),
      fs = require('fs'),
      count = 0;

  app.get('/', >> handler); //@\label{lst:fluxional-streamtohandler}@
  app.listen(8080);

flx handler
-> reply [res]
  function handler(req, res) {
    fs.readFile(__filename, -> reply); //@\label{lst:fluxional-readfile}@
  }

flx reply & grp_res {count, template}
-> null
  function reply(error, data) {
    count += 1; //@\label{lst:fluxional-counter}@
    res.send(err || template(count, data)); //@\label{lst:fluxional-ressend}@
  }
\end{code}

The chain of functions from listing \ref{lst:source} is expressed in the fluxional language in listing \ref{lst:fluxional}.
The fluxion \texttt{handler} doesn't have any dependencies, so it can be executed in a parallel event-loop.
The fluxions \texttt{main} and \texttt{reply} belong to the group \texttt{grp\_res}, indicating their dependency over the variable \texttt{res}.
The group name is arbitrarily chosen by the compiler.
All the fluxions inside a group are executed sequentially on the same event-loop, to protect against concurrent accesses.

The variable \texttt{res} is created and consumed within a chain of \textit{post} stream.
Therefore, it is exclusive to one request and cannot be propagated to another request.
It doesn't prevent the whole group from being replicated.
However, the fluxion \texttt{reply} depends on the variable \texttt{count} created upstream the \textit{start} stream, which prevents this replication.
If it did not rely on this state, the group \texttt{grp\_res} would be stateless, and could be replicated to cope with the incoming traffic.

This execution model allows to parallelize the execution of an application as a pipeline, as with the fluxion \texttt{handler}.
And some parts are replicated, as could be the group \texttt{grp\_res}.
This parallelization improves the scalability of the application.
Indeed, as a fluxion contains its state and expresses its dependencies, it can be migrated.
It allows to adapt the number of fluxions per core to adjust the resource usage in function of the desired throughput.

Our goal, as described in the introduction, is not to propose a new high-level language but to automate the architectural shift.
We present the compiler to automate this architectural shift in the next section.
\section{Fluxionnal compiler} \label{section:compiler}

The source languages we focus on should offer higher-order functions and be implemented as an event-loop with a global memory.
Javascript is such a language and is often implemented on top of an event-loop, like in \textit{Node.js}.
We developed a compiler that transforms a \textit{Node.js} application into a fluxional application compliant with the execution model described in section \ref{section:model}.
Our compiler uses the \textit{estools}\ftnt{https://github.com/estools} suite to parse, manipulate and generate source code from Abstract Syntax Tree (AST).
And it is tailored for -- but not limited to -- web applications using \textit{Express}\ftnt{http://expressjs.com/}, the most used \textit{Node.js} web framework.

\begin{figure}
  \includegraphics[width=\linewidth]{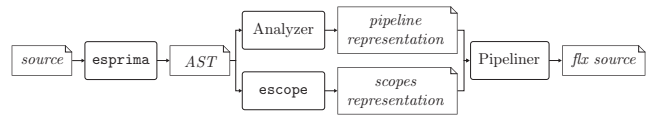}
  \caption{Compilation chain}
  \label{fig:compilation}
\end{figure}

The chain of compilation is described in figure \ref{fig:compilation}.
The compiler extracts an AST from the source with \texttt{esprima}.
From this AST, the \textit{Analyzer} step identifies the limits of the different application parts and how they relate to form a pipeline.
This first step outputs a pipeline representation of the application.
Section \ref{section:compiler:analyzer} explains this first compilation step.
In the pipeline representation, the stages are not yet independent and encapsulated into fluxions.
From the AST, \texttt{escope} produces a representation of the memory scopes.
The \textit{Pipeliner} step analyzes the pipeline representation and the scopes representation to distribute the shared memory into independent groups of fluxions.
Section \ref{section:compiler:pipeliner} explains this second compilation step.

\subsection{Analyzer step} \label{section:compiler:analyzer}

The limit between two application parts is defined by a rupture point.
The analyzer identifies these rupture points, and outputs a representation of the application in a pipeline form.
Application parts are the stages, and rupture points are the message streams of this pipeline.

\subsubsection{Rupture points} \label{section:compiler:analyzer:rupture}

\begin{figure}[h!]
  \includegraphics[width=\linewidth]{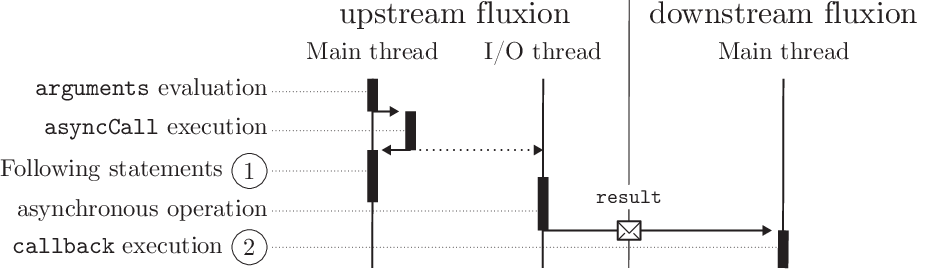}
  \begin{code}
asyncCall(arguments, function callback(result){ //@\circled{2}@ });
// Following statements //@\circled{1}@
  \end{code}
  \caption{Rupture point interface}
  \label{fig:basicrp}
\end{figure}

A rupture point is a call of a loosely coupled function.
It is an asynchronous call without subsequent synchronization with the caller.
In \textit{Node.js}, I/O operations are asynchronous functions and indicate rupture points between two application parts.
Figure \ref{fig:basicrp} shows a code example of a rupture point with the illustration of the execution of the two application parts isolated into fluxions.
The two application parts are the caller of the asynchronous function call on one hand, and the callback provided to the asynchronous function call on the other hand.

A callback is a function passed as a parameter to a function call.
It is invoked by the callee to continue the execution with data not available in the caller context.
There are three kinds of callbacks, but only two are asynchronous: listeners and continuations.
The two corresponding types of rupture points are \textit{start} and \textit{post}.

\textbf{Start rupture points} (listeners) are on the border between the application and the outside, continuously receiving incoming user requests.
An example of a start rupture point is in listing \ref{lst:source}, between the call to \texttt{app.get()}, and its listener \texttt{handler}.
These rupture points indicate the input of a data stream in the program, and the beginning of a chain of fluxions to process this stream.

\textbf{Post rupture points} (continuations) represent a continuity in the execution flow after an asynchronous operation yielding a unique result, such as reading a file, or a database.
An example of a post rupture points is in listing \ref{lst:source}, between the call to \texttt{fs.readFile()}, and its continuation \texttt{reply}.

\subsubsection{Detection}

The compiler uses a list of common asynchronous callees, like the \texttt{express} and file system methods.
This list can be augmented to match asynchronous callees individually for any application.
To identify the callee, the analyzer walks the AST to find a call expression matching this list.

After the identification of the callee, the callback needs to be identified as well, to be encapsulated in the downstream fluxion.
For each asynchronous call detected, the compiler tests if one of the arguments is of type \texttt{function}.
Some callback functions are declared \textit{in situ}, and are trivially detected.
For variable identifiers, and other expressions, the analyzer tries to detect their type.
The analyzer walks back the AST to track their assignations and modifications, so as to determine their last value.

\subsection{Pipeliner step} \label{section:compiler:pipeliner}

\begin{figure}[h!]
  \includegraphics[width=\linewidth]{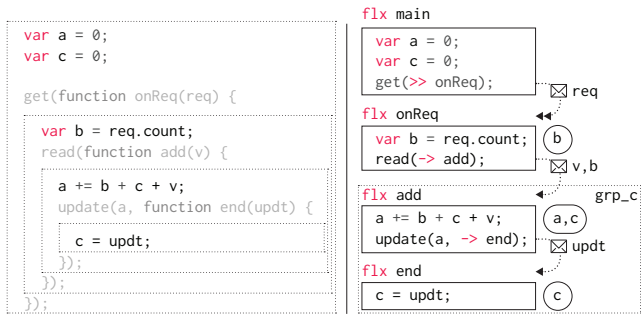}
  \caption{Variable management from Javascript to the high-level fluxionnal language}
  \label{fig:states}
\end{figure}

A rupture point eventually breaks the chain of scopes between the upstream and downstream fluxion.
The closure in the downstream fluxion cannot access the scope in the upstream fluxion as expected.
The pipeliner step replaces the need for this closure, allowing application parts to rely only on independent memory stores and message passing.
It determines the distribution using the scope representation, which represents the variables' dependencies between application parts.
Depending on this representation, the compiler can replace the broken closures in three different ways.
We present these three alternatives in figure \ref{fig:states}.

\paragraph{Scope}
If a variable is modified inside only one application part in the current \textit{post} chain, then the pipeliner adds it to the context of its fluxion.

In figure \ref{fig:states}, the variable \texttt{a} is updated in the function \texttt{add}.
The pipeliner step stores this variable in the context of the fluxion \texttt{add}.

\paragraph{Stream}
If a modified variable is read by downstream application parts, then the pipeliner makes the upstream fluxion add this variable to the message stream to be sent to the downstream fluxions.
It is impossible to send variables to upstream flux\-ions, without causing inconsistencies.
If the fluxion retro propagates the variable for an upstream fluxion to read, the upstream fluxion might use the old version while the new version is on its way.

In figure \ref{fig:states}, the variable \texttt{b} is set in the function \texttt{onReq}, and read in the function \texttt{add}.
The pipeliner step makes the fluxion \texttt{onReq} send the updated variable \texttt{b}, in addition to the variable \texttt{v}, in the message sent to the fluxion \texttt{add}.

Exceptionally, if a variable is defined inside a \textit{post} chain, like \texttt{b}, then this variable can be streamed inside this \textit{post} chain without restriction on the order of modification and read.
Indeed, the execution of the upstream fluxion for the current \textit{post} chain is assured to end before the execution of the downstream fluxion.
Therefore, no reading of the variable by the upstream fluxion happens after the modification by the downstream fluxion.

\paragraph{Share}
If a variable is needed for modification by several application parts, or is read by an upstream application part, then it needs to be synchronized between the fluxions.
To respect the semantics of the source application, we cannot tolerate inconsistencies.
Therefore, the pipeliner groups all the fluxions sharing this variable with the same tag.
And it adds this variable to the contexts of each fluxions.

In figure \ref{fig:states}, the variable \texttt{c} is set in the function \texttt{end}, and read in the function \texttt{add}.
As the fluxion \texttt{add} is upstream of \texttt{end}, the pipeliner step groups the fluxion \texttt{add} and \texttt{end} with the tag \texttt{grp\_c} to allow the two fluxions to share this variable.
\section{Real test case} \label{section:evaluation}

This section presents a test of the compiler on a real application, gifsockets-server\ftnt{https://github.com/twolfson/gifsockets-server}.
This test proves the possibility for an application to be compiled into a network of independent parts.
It shows the current limitations of this isolation and the modifications needed on the application to circumvent them.
This section then presents future works.

\begin{code}[js, caption={Simplified version of gifsockets-server},label={lst:gifsocket}]
var express = require('express'),
    app = express(),
    routes = require('gifsockets-middleware'), //@\label{lst:gifsocket:gif-mw}@
    getRawBody = require('raw-body');

function bodyParser(limit) { //@\label{lst:gifsocket:bodyParser}@
  return function saveBody(req, res, next) { //@\label{lst:gifsocket:saveBody}@
    getRawBody(req, { //@\label{lst:gifsocket:getRawBody}@
      expected: req.headers['content-length'],
      limit: limit
    }, function (err, buffer) { //@\label{lst:gifsocket:callback}@
      req.body = buffer;
      next(); //@\label{lst:gifsocket:next}@
    });
  };
}

app.post('/image/text', bodyParser(1 * 1024 * 1024), routes.writeTextToImages); //@\label{lst:gifsocket:app.post}@
app.listen(8000);
\end{code}

This application, simplified in listing \ref{lst:gifsocket}, is a real-time chat using gif-based communication channels.
It was selected in a previous work \cite{Brodu2015} from the \texttt{npm} registry because it depends on \texttt{express}, it is tested, working, and simple enough to illustrate this evaluation.
The server transforms the received text into a gif frame, and pushes it back to a never-ending gif to be displayed on the client.

On line \ref{lst:gifsocket:app.post}, the application registers two functions to process the requests received on the url \texttt{/image/text}.
The closure \texttt{saveBody}, line \ref{lst:gifsocket:saveBody}, returned by \texttt{bodyParser}, line \ref{lst:gifsocket:bodyParser}, and the method \texttt{routes.write\-Text\-To\-Images} from the external module \texttt{gifsockets-middleware}, line \ref{lst:gifsocket:gif-mw}.
The closure \texttt{saveBody} calls the asynchronous function \texttt{getRawBody} to get the request body.
Its callback handles the errors, and calls \texttt{next} to continue processing the request with the next function, \texttt{routes.write\-Text\-To\-Images}.

\subsection{Compilation}

We compile this application with the compiler detailed in section \ref{section:compiler}.
Listing \ref{lst:flx-gifsocket} presents the compilation result.
The function call \texttt{app.post}, line \ref{lst:gifsocket:app.post}, is a rupture point.
However, its callbacks, \texttt{bodyParser} and \texttt{routes.write\-Text\-To\-Images} are evaluated as functions only at runtime.
For this reason, the compiler ignores this rupture point, to avoid interfering with the evaluation.

\begin{code}[flx, caption={Compilation result of gifsockets-server},label={lst:flx-gifsocket}]
flx main & express {req}
>> anonymous_1000 [req, next]
  var express = require('express'),
      app = express(),
      routes = require('gifsockets-middleware'), //@\label{lst:flx-gifsocket:gif-mw}@
      getRawBody = require('raw-body');

  function bodyParser(limit) { //@\label{lst:flx-gifsocket:bodyParser}@
    return function saveBody(req, res, next) { //@\label{lst:flx-gifsocket:saveBody}@
      getRawBody(req, { //@\label{lst:flx-gifsocket:getRawBody}@
        expected: req.headers['content-length'], //@\label{lst:flx-gifsocket:req.headers}@
        limit: limit
      }, >> anonymous_1000);
    };
  }

  app.post('/image/text', bodyParser(1 * 1024 * 1024), routes.writeTextToImages); //@\label{lst:flx-gifsocket:app.post}@
  app.listen(8000);

flx anonymous_1000
-> null
  function (err, buffer) { //@\label{lst:flx-gifsocket:callback}@
    req.body = buffer; //@\label{lst:flx-gifsocket:buffer}@
    next(); //@\label{lst:flx-gifsocket:next}@
  }
\end{code}

The compiler detects a rupture point : the function \texttt{get\-Raw\-Body} and its anonymous callback, line \ref{lst:gifsocket:callback}.
It encapsulates this callback in a fluxion named \texttt{anonymous\_\-1000}.
The callback is replaced with a stream placeholder to send the message stream to this downstream fluxion.
The variables \texttt{req} and \texttt{next} are appended to this message stream, to propagate their value from the \texttt{main} fluxion to the \texttt{anonymous\_\-1000} fluxion.

When \texttt{anonymous\_\-1000} is not isolated from the \texttt{main} fluxion, as if they belong to the same group, the compilation result works as expected.
The variables used in the fluxion, \texttt{req} and \texttt{next}, are still shared between the two fluxions.
Our goal is to isolate the two fluxions, to be able to safely parallelize their executions.

\eject

\subsection{Isolation}

In listing \ref{lst:flx-gifsocket}, the fluxion \texttt{anonymous\_1000} modifies the object \texttt{req}, line \ref{lst:flx-gifsocket:buffer}, to store the text of the received request, and it calls \texttt{next} to continue the execution, line \ref{lst:flx-gifsocket:next}.
These operations produce side-effects that should propagate in the whole application, but the isolation prevents this propagation.
Isolating the fluxion \texttt{anonymous\_1000} produces runtime exceptions.
We detail in the next paragraph, how we handle this situation to allow the application to be parallelized.

\subsubsection{Variable \texttt{req}}

The variable \texttt{req} is read in fluxion \texttt{main}, lines \ref{lst:flx-gifsocket:getRawBody} and \ref{lst:flx-gifsocket:req.headers}.
Then its property \texttt{body} is associated to \texttt{buffer} in fluxion \texttt{anonymous\_1000}, line \ref{lst:flx-gifsocket:buffer}.
The compiler is unable to identify further usages of this variable.
However, the side effect resulting from this association impacts a variable in the scope of the next callback, \texttt{routes.writeTextToImages}.
We modified the application to explicitly propagate this side-effect to the next callback through the function \texttt{next}.
We explain further modification of this function in the next paragraph.

\subsubsection{Closure \texttt{next}}

The function \texttt{next} is a closure provided by the \texttt{express} \texttt{Router} to continue the execution with the next function to handle the client request.
Because it indirectly relies on the variable \texttt{req}, it is impossible to isolate its execution with the \texttt{anonymous\_\-1000} fluxion.
Instead, we modify \texttt{express}, so as to be compatible with the fluxionnal execution model.
We explain the modifications below.

\begin{code}[flx, caption={Simplified modification on the compiled result},label={lst:mflx-gifsocket}]
flx anonymous_1000
-> express_dispatcher
  function (err, buffer) { //@\label{lst:mflx-gifsocket:callback}@
    req.body = buffer; //@\label{lst:mflx-gifsocket:buffer}@
    next_placeholder(req, -> express_dispatcher); //@\label{lst:mflx-gifsocket:next-placeholder}@
  }

flx express_dispatcher & express {req} //@\label{lst:mflx-gifsocket:express-dispatcher}@
-> null
  function (modified_req) {
    merge(req, modified_req);
    next(); //@\label{lst:mflx-gifsocket:next}@
  }
\end{code}

In listing \ref{lst:gifsocket}, the function \texttt{next} is a continuation allowing the anonymous callback, line \ref{lst:gifsocket:callback}, to call the next function to handle the request.
To isolate the anonymous callback into \texttt{anonymous\_\-1000}, \texttt{next} is replaced by a rupture point.
This replacement is illustrated in listing \ref{lst:mflx-gifsocket}.
The \texttt{express} \texttt{Router} registers a fluxion named \texttt{express\_\-dispatcher}, line \ref{lst:mflx-gifsocket:express-dispatcher}, to continue the execution after the fluxion \texttt{anonymous\_\-1000}.
This fluxion is in the same group \texttt{express} as the \texttt{main} fluxion, hence it has access to the original variable \texttt{req}, and to the original function \texttt{next}.
The call to the original \texttt{next} function is replaced by a placeholder to push the stream to the fluxion \texttt{express\_\-dispatcher}, line \ref{lst:mflx-gifsocket:next-placeholder}.
The fluxion \texttt{express\_\-dispatcher} receives the stream from the upstream fluxion \texttt{anonymous\_\-1000}, merges back the modification in the variable \texttt{req} to propagate the side effects, and finally calls the original function \texttt{next} to continue the execution, line \ref{lst:mflx-gifsocket:next}.

After the modifications detailed above, the server works as expected.
The isolated fluxion correctly receives, and returns its serialized messages.
The client successfully receives a gif frame containing the text.

\subsection{Future works}

We intend to implement the compilation process presented into the runtime.
A just-in-time compiler would allow to identify callbacks dynamically evaluated, and to analyze the memory to identify side-effects propagations instead of relying only on the source code.
Moreover, this memory analysis would allow the closure serialization required to compile application using higher-order functions.
\section{Related Works} \label{section:related}

Splitting a task into independent parts goes back to the Actor's model, functional programming \cite{Hughes1989} and the following works on Data\-Flow leading to Flow-based Programming (FBP) and Functional Reactive program\-ming (FRP) \cite{Elliott1997}.
Both FBP and FRP, recently got some attention in the Javascript community with \textit{NoFlo}\ftnt{http://noflojs.org/}, \textit{Bacon.js}\ftnt{https://baconjs.github.io/} and \textit{react}\ftnt{https://facebook.github.io/react/}.

The execution model we presented in section \ref{section:model}, is inspired by works on scalability for very large systems, like the Staged Event-Driven Architecture (SEDA) by Matt Welsh \cite{Welsh2000} and by the MapReduce architecture \cite{Dean2008}.
It also drew its inspiration from more recent work following SEDA like Spark \cite{Zaharia2012}, MillWheel \cite{Akidau2013}, Naiad \cite{McSherry} and Storm \cite{Toshniwal2014}.
The first part of our work stands upon these thorough studies.
However, we believe that it is difficult for most developers to distribute the state of an application.
This belief motivated us to propose a compiler from an imperative programming model to these more scalable, distributed execution engines.

The transformation of an imperative programming model to be executed onto a parallel execution engine was recently addressed by Fernandez \textit{et. al.} \cite{Fernandez2014a}.
However, as in similar works \cite{Power2010}, it requires annotations from developers, therefore partially conserves the disruption with the feature-based development.
Our approach discards the need for annotations, thus targets a broader range of developers than only ones experienced with parallel development.

A great body of work focuses on parallelizing sequential programs \cite{Banerjee2013,Li2012,Matsakis2012a,Radoi2014}.
Because of the synchronous execution of a sequential program, the speedup of parallelization is inherently limited \cite{Amdahl1967,Gunther2008}.
On the other hand, our approach is based on an asynchronous programming model.
Hence the attainable speedup is not limited by the main synchronous thread of execution.
\section{Conclusion} \label{section:conclusion}

In this paper, we presented our work on a high-level language allowing to represent a web application as a network of independent parts communicating by message streams.
We presented a compiler to transform a \textit{Node.js} web application into this high-level representation.
To identify two independent parts, the compiler spots rupture points in the application, possibly leading to memory isolation and thus, parallelism.
We presented an example of a compiled application to show the limits of this approach.
The parallelism of this approach allows code-mobility which may lead to a better scalability.
We believe it can enable the scalability required by highly concurrent web applications without discarding the familiar, feature-driven programming models.

\printbibliography[]

\end{document}